\documentclass[]{spie}  

\usepackage{amsmath,amsfonts,amssymb}
\usepackage{graphicx}
\usepackage[colorlinks=true, allcolors=blue]{hyperref}

\title{Design and simulation of a photonic lantern-inspired astrophotonic chip for spectral sensing}

\author[a]{Avi Patel}
\author[a]{Kevin A. Bundy}
\author[a]{Aditya Sengupta}
\author[a]{Matthew C. DeMartino}
\author[a]{Anna Gagnebin}
\author[a]{Emiel Por}
\author[b]{Majid Mohammad}
\author[b]{Michael Arena}
\author[b]{Aled Cuda}
\author[b]{Kiana Ejercito}
\author[c]{Stephen Eikenberry}
\author[b]{Ben Mazin}
\author[a]{Holger Schmidt}

\affil[a]{Department of Astronomy \& Astrophysics, University of California, Santa Cruz, CA 95064,
USA}
\affil[b]{Department of Physics, University of California, Santa Barbara, CA 93106,
USA}
\affil[c]{CREOL, The College of Optics \& Photonics, University of Central Florida, Orlando, FL
32816, USA}

\authorinfo{Further author information: (Send correspondence to A.P.)\\A.P.: E-mail: avpapate@ucsc.edu}

\begin{document} 
\maketitle

\begin{abstract}
Compact astrophotonic sensors can trade general-purpose spectral coverage for task-specific wavelength discrimination in a small integrated footprint. Developed with the Mazin Lab at UC Santa Barbara, this chip couples a single-mode input into a multimode interference region and seven-port fanout, producing wavelength-dependent output power fingerprints for spectral retrieval. Using ANSYS Lumerical FDTD simulations in silicon nitride, we compare symmetric and staggered-release geometries and quantify throughput, wavelength-dependent changes in the seven-port output distribution, and sensitivity near a 745 nm design point. We optimize over the tested geometric parameter space and identify designs with improved throughput and wavelength discrimination, including increased throughput-weighted Fisher information relative to the baseline. These results suggest that lantern-inspired integrated photonics can enable compact, task-specific spectral sensors for astrophotonic applications.
\end{abstract}

\keywords{Astrophotonics, photonic lantern, on-chip spectral sensing, multimode interference}

\section{INTRODUCTION}
\label{sec:introduction}  

Spectroscopy is central to astronomical instrumentation, chemical and
biological sensing, telecommunications, and environmental monitoring. In
astronomy, spectral measurements provide access to the composition,
temperature, velocity, and magnetic environment of otherwise unresolved
sources. Conventional spectrometers generally separate wavelengths using a
dispersive element, such as a diffraction grating or prism, and form a
spatially resolved spectrum on a detector. Although this architecture is
mature and capable of high resolving power, its performance is commonly tied
to optical path length, aperture size, and detector format. These constraints
motivate integrated and computational spectrometers that replace bulk
dispersion with a calibrated, wavelength-dependent optical response
\cite{amin2024mmi}.

Reconstructive spectrometers exploit the fact that coherent propagation
through a multimode structure produces a deterministic spatial pattern. An
input field coupled into a multimode waveguide excites a superposition of
modes with distinct, wavelength-dependent propagation constants. The relative
phases accumulated by these modes therefore vary with both propagation
distance and wavelength, causing the output intensity distribution to act as a
spectral fingerprint. Recent work has demonstrated this principle using
multimode-interference (MMI) waveguides whose internal scattering patterns are
recorded by a camera and decoded computationally. Amin \textit{et al.}
demonstrated visible and near-infrared operation, including 0.05-nm spectral
resolution near 800~nm and multiple spectrometers integrated on one chip
\cite{amin2024mmi}. These results establish multimode interference as a
powerful mechanism for compact spectral encoding, but imaging the full
propagation pattern requires a camera, out-of-plane scattering, and a
calibration or reconstruction stage.

Photonic lanterns provide a complementary method of accessing information
contained in a multimode field. A conventional photonic lantern adiabatically
transforms a multimode waveguide into an array of single-mode waveguides,
providing a low-loss interface when the numbers of supported spatial degrees
of freedom are appropriately matched \cite{birks2015photoniclantern}. Beyond
their original role as mode converters, photonic lanterns can be viewed as
modal analyzers: coherent changes in the field at the multimode end alter the
relative complex amplitudes, and therefore the detected intensities, at the
single-mode outputs. The resulting vector of output powers is a compact
measurement of the incident field that can be acquired using a small detector
array rather than an image of the complete multimode propagation region. This
property has motivated photonic-lantern concepts for focal-plane wavefront
sensing, high-contrast imaging, and spectroastrometry
\cite{lin2022plwfs,kim2024potential}.

The spectral dependence of this modal mapping is particularly important for
astrophotonic instruments. In photonic-lantern spectroastrometry, for example,
the relative intensities of the single-mode outputs contain information about
wavelength-dependent shifts in the centroid of an astronomical source.
Numerical studies of a six-port lantern showed that suitably designed lantern
responses can provide sensitivity to two-dimensional spectroastrometric
signals, while also highlighting the effects of photon noise, residual
wavefront error, and chromaticity \cite{kim2024potential}. Kim \textit{et al.}
subsequently characterized a three-port lantern experimentally on the Subaru
Coronagraphic Extreme Adaptive Optics testbed. Their measurements reproduced
the predicted wavelength-dependent, approximately sinusoidal astrometric
sensitivity and demonstrated a method for constructing wavelength-dependent
transfer-matrix models from measured coupling maps
\cite{kim2024spectral}. This work emphasizes both the information available in
the relative output powers and the need to characterize and calibrate their
chromatic response.

Recent laboratory and reconstruction studies have extended this approach
toward larger lantern systems and direct spectral inference. Sengupta
\textit{et al.} used digital off-axis holography to measure the principal
modes of individual lanterns in an integrated array of seven 19-port photonic
lanterns and compared the measured modes with simulations to assess
manufacturing variation \cite{sengupta2026holography}. Their results reinforce
the importance of empirical transfer-function calibration when a lantern is
used as a quantitative sensor. In parallel, DeMartino \textit{et al.} have
investigated neural-network reconstruction of spectra from photonic-lantern
outputs, while Gagnebin \textit{et al.} have studied the resulting spectral
reconstruction performance
\cite{demartino2026neural,gagnebin2026performance}. Together, these efforts
show that a photonic lantern can serve not only as a mode converter but also
as the calibrated optical encoder in a computational spectrometer.

The same chromatic behavior that must be calibrated in a spatial sensor can
also be intentionally engineered for wavelength discrimination. If a
multimode section and its fanout are designed so that a small wavelength
change produces a large, nondegenerate redistribution of power among the
single-mode outputs, the lantern becomes a compact spectral encoder. Unlike a
camera-based MMI spectrometer, such a device samples the interference pattern
at a finite number of guided output channels. Its performance is consequently
determined not only by the wavelength dependence of the normalized output
vector, but also by the total transmitted power and by the statistical
information carried by the detected photons. This creates a geometry-design
problem in which spectral sensitivity, throughput, modal purity, device
length, and fabrication constraints must be considered together.

In this work, we investigate a planar seven-output silicon-nitride photonic
lantern designed for operation near 745~nm. A single-mode input excites a
shared multimode region, after which an asymmetric sequential-release fanout
maps the evolving interference pattern into seven separated output
waveguides. Three-dimensional finite-difference time-domain simulations are
used to optimize the longitudinal extent of the multimode and fanout regions
and to evaluate the wavelength-dependent output fingerprint. The response is
quantified using the Euclidean slope of the normalized seven-port intensity
vector and the corresponding multinomial Fisher information, while total
throughput is retained as an independent performance measure. We further
perform a time-convergence study of the refined geometry and examine both the
output-plane field profiles and the longitudinal multimode evolution.

The resulting design produces distinct output fingerprints over the
741--749-nm interval while maintaining substantial simulated throughput. More
broadly, the study demonstrates a route toward camera-free spectral
discrimination in which wavelength is encoded directly into a small number of
single-mode output intensities. The present analysis focuses on device-level
spectral sensitivity and numerical convergence; experimental calibration,
fabrication tolerances, and reconstruction of unknown or broadband spectra
remain subjects for subsequent work.

\section{Photonic Lantern Chip Design}

\subsection{PL-Chip Design Parameterization}
\label{sec:pl_chip_design}

\begin{figure}
    \centering
    \includegraphics[width=1\linewidth]{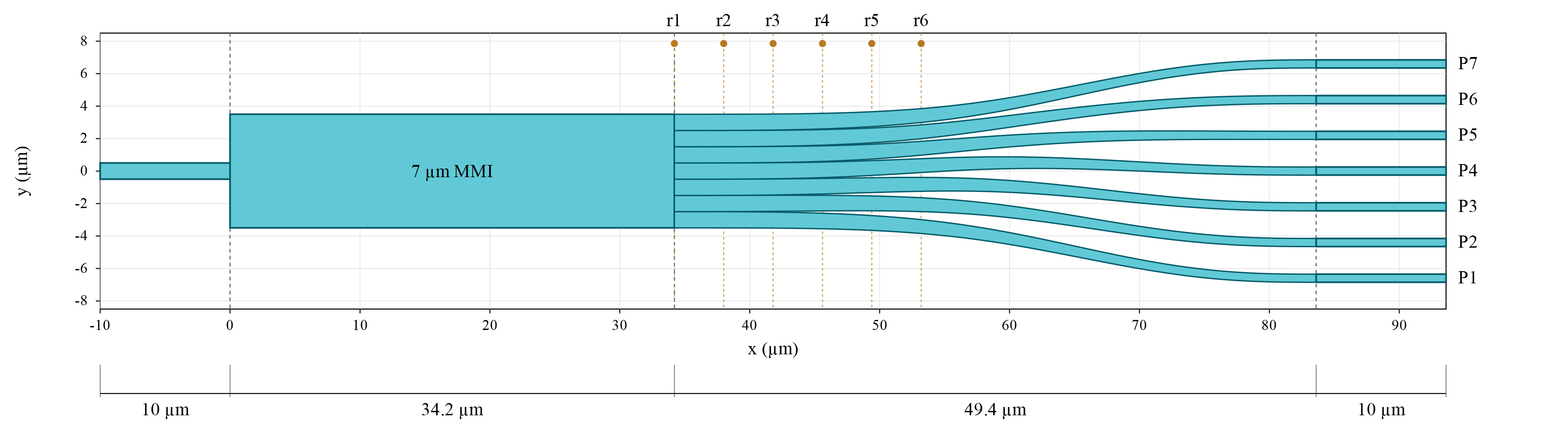}
    \caption{Parameterized geometry of the seven-port photonic-lantern chip. A $1~\mu\mathrm{m}$-wide input waveguide feeds a $7~\mu\mathrm{m}$-wide MMI region, followed by a staggered-release fanout with release positions $r_1$--$r_6$. The branches taper to $0.5~\mu\mathrm{m}$-wide output waveguides on a $2.2~\mu\mathrm{m}$ pitch. All dimensions are in micrometers.}
    \label{fig:PL_chip}
\end{figure}

Figure~\ref{fig:PL_chip} shows the parameterized seven-port photonic-lantern (PL) chip considered in this work. The device comprises four longitudinal sections: a fundamental-mode-excited access waveguide, a multimode-interference (MMI) region, a staggered-release fanout with simultaneous width reduction, and seven isolated output waveguides. The coordinate origin is placed at the junction between the access waveguide and the MMI region.

The waveguides are formed from a \(200~\mathrm{nm}\)-thick Si\(_3\)N\(_4\) strip core with a SiO\(_2\) upper cladding and a substrate refractive index of 1.70. The structure was designed for TE-like excitation near \(745~\mathrm{nm}\) and evaluated over the \(720\)--\(770~\mathrm{nm}\) wavelength range. Light is launched into the fundamental TE-like mode of a centered, \(1~\mu\mathrm{m}\)-wide access waveguide. 

At \(x=0\), the access waveguide expands directly into a \(7~\mu\mathrm{m}\)-wide MMI region. This nonadiabatic transition provides a compact and reproducible projection of the input field onto several modes of the wider section. The wavelength-dependent phase accumulation among these modes establishes the interference field that is subsequently sampled by the
seven output branches. The MMI width also accommodates seven contiguous \(1~\mu\mathrm{m}\)-wide branch mouths at the beginning of the fanout.

The longitudinal geometry was parameterized using a scale factor \(s_x\) applied to the MMI and fanout coordinates of the initial design. For the geometry shown in Fig.~\ref{fig:PL_chip}, \(s_x=0.76\), giving $x_{\mathrm{MMI}} = 45s_x = 34.2~\mu\mathrm{m}$ for the end of the uniform MMI region and $x_{\mathrm{F}} = 110s_x = 83.6~\mu\mathrm{m}$ for the end of the fanout. The resulting fanout length is therefore $L_{\mathrm{F}} = x_{\mathrm{F}}-x_{\mathrm{MMI}}= 49.4~\mu\mathrm{m}$. The selection of \(s_x=0.76\) is discussed separately with the refined simulation results.

At \(x=x_{\mathrm{MMI}}\), the seven branch mouths each have a width of \(1~\mu\mathrm{m}\) and tile the \(7~\mu\mathrm{m}\)-wide MMI region without intervening gaps. Rather than separating all branches at the same longitudinal position, the six boundaries between adjacent branches are released sequentially. Their release coordinates are given by, 
\begin{equation}
x_{r,j}
= s_x\left[45+5(j-1)\right],
\qquad j=1,\ldots,6,
\end{equation}
which gives,
$x_{r,j}
=
\left\{
34.2,\,
38.0,\,
41.8,\,
45.6,\,
49.4,\,
53.2
\right\}
~\mu\mathrm{m}$.
This staggered release progressively breaks the transverse symmetry of the device and distributes the geometric transition over the fanout length. 

Each separation is described using the quintic minimum-jerk function,
\begin{equation}
S(u)=10u^{3}-15u^{4}+6u^{5},
\end{equation}
where the normalized coordinate associated with the \(j\)-th release is,
\begin{equation}
u_j(x)
=
\operatorname{clip}
\left(
\frac{x-x_{r,j}}{x_{\mathrm{F}}-x_{r,j}},
0,1
\right).
\end{equation}

The local center-to-center separation across boundary \(j\) is parameterized
as

\begin{equation}
d_j(x)
=
w_{\mathrm{m}}
+
\left(P-w_{\mathrm{m}}\right)
S\!\left[u_j(x)\right],
\end{equation}

where \(w_{\mathrm{m}}=1~\mu\mathrm{m}\) is the branch-mouth width and
\(P=2.2~\mu\mathrm{m}\) is the final output pitch. The branches therefore
remain packed before their respective release positions and smoothly approach
the required \(2.2~\mu\mathrm{m}\) separation toward the output plane.

The center position of branch \(p\), with \(p=1,\ldots,7\), can be written as

\begin{equation}
y_p(x)
=
\sum_{j=1}^{p-1} d_j(x)
-
\frac{1}{2}
\sum_{j=1}^{6} d_j(x).
\end{equation}

This construction maintains the branch bundle about \(y=0\) throughout the
transition.

The branch widths are simultaneously reduced from \(1~\mu\mathrm{m}\) at the
MMI interface to \(0.5~\mu\mathrm{m}\) at the end of the fanout. The width
profile is

\begin{equation}
w(x)
=
w_{\mathrm{m}}
+
\left(w_{\mathrm{o}}-w_{\mathrm{m}}\right)
S\!\left[
\frac{x-x_{\mathrm{MMI}}}
{x_{\mathrm{F}}-x_{\mathrm{MMI}}}
\right],
\end{equation}

where \(w_{\mathrm{o}}=0.5~\mu\mathrm{m}\) is the final output-waveguide
width. Because the first derivative of \(S(u)\) vanishes at both endpoints,
the branch widths and trajectories join the packed input and straight output
sections without slope discontinuities.

At the end of the fanout, the output centers are located at

\begin{equation}
y_p=(p-4)P,
\qquad p=1,\ldots,7,
\end{equation}
corresponding to
$
y_p
=
\left\{
-6.6,\,
-4.4,\,
-2.2,\,
0,\,
2.2,\,
4.4,\,
6.6
\right\}
~\mu\mathrm{m}$.

Each branch then continues through a \(10~\mu\mathrm{m}\)-long straight
output waveguide with a width of \(0.5~\mu\mathrm{m}\). The final
\(2.2~\mu\mathrm{m}\) pitch provides an edge-to-edge separation of
\(1.7~\mu\mathrm{m}\) between neighboring output cores.

\begin{table}[t]
\centering
\caption{Geometric parameters of the seven-port PL chip.}
\label{tab:pl_design_parameters}
\begin{tabular}{l c}
\hline
\textbf{Parameter} & \textbf{Final value} \\
\hline
Si\(_3\)N\(_4\) core thickness
    & \(0.20~\mu\mathrm{m}\) \\
Access-waveguide width
    & \(1.0~\mu\mathrm{m}\) \\
Access-waveguide length
    & \(10~\mu\mathrm{m}\) \\
MMI width
    & \(7.0~\mu\mathrm{m}\) \\
MMI length
    & \(34.2~\mu\mathrm{m}\) \\
Longitudinal scale factor, \(s_x\)
    & \(0.76\) \\
Fanout length
    & \(49.4~\mu\mathrm{m}\) \\
Branch-mouth width
    & \(1.0~\mu\mathrm{m}\) \\
Final output width
    & \(0.5~\mu\mathrm{m}\) \\
Final output pitch
    & \(2.2~\mu\mathrm{m}\) \\
Release-point spacing
    & \(3.8~\mu\mathrm{m}\) \\
Straight output-tail length
    & \(10~\mu\mathrm{m}\) \\
Number of output ports
    & \(7\) \\
\hline
\end{tabular}
\end{table}
Quantitative transmission, spectral sensitivity, and the segment-by-segment power budget are presented later with the refined simulation results.

\section{Simulations}

\subsection{Optimization metrics and sweep procedure}
\label{subsec:optimization_metrics}

All electromagnetic simulations were performed using the three-dimensional
finite-difference time-domain solver in Ansys Lumerical FDTD, Release 2025 R2
\cite{ansyslumerical2025r2}. The device was excited with the fundamental TE
input mode, and the powers transmitted through the seven output-monitor
planes were recorded as functions of wavelength.

The geometry was optimized for wavelength-dependent redistribution of power
among the output ports rather than for throughput alone. At each wavelength,
the simulated output powers \(P_i\) were converted into a normalized intensity
fingerprint,
\begin{equation}
    p_i(\lambda)
    =
    \frac{P_i(\lambda)}
    {\displaystyle\sum_{j=1}^{7}P_j(\lambda)}
    \label{eq:normalized_fingerprint}
\end{equation}
Two local metrics were used to quantify the wavelength dependence of this
fingerprint. The spectral sensitivity was defined as
\begin{equation}
    S_{\lambda}
    =
    \left\|
        \frac{\partial\mathbf{p}}{\partial\lambda}
    \right\|_2
    \label{eq:spectral_sensitivity}
\end{equation}
where \(\mathbf{p}=[p_1,\ldots,p_7]^{\mathrm{T}}\). This quantity measures the
overall rate at which the normalized seven-port pattern changes with
wavelength, treating changes in all ports equally.

The corresponding multinomial Fisher information was calculated as
\begin{equation}
    \mathcal{I}_{\lambda}
    =
    \sum_{i=1}^{7}
    \frac{1}{p_i}
    \left(
        \frac{\partial p_i}{\partial\lambda}
    \right)^2
    \label{eq:fisher_information}
\end{equation}
This expression follows from an ideal photon-counting model in which each
detected photon is assigned to one of the seven output ports with probability
\(p_i\). For a fixed total of \(N\) detected photons, the seven port counts are
jointly multinomial: an increase in the count at one port necessarily reduces
the fraction assigned to the others. Under independent-photon, shot-noise-
limited detection, \(\mathcal{I}_{\lambda}\) is the Fisher information per
detected photon and \(N\mathcal{I}_{\lambda}\) is the information in the full
set of counts. This model does not include detector read noise, dark counts,
background light, calibration error, or uncertainty in the total input power;
the Fisher information is therefore used here as an idealized sensitivity
metric rather than as a complete detector-noise model.

Because both \(S_{\lambda}\) and \(\mathcal{I}_{\lambda}\) are calculated from
normalized fractions, they do not account for how much optical power reaches
the outputs. A geometry could therefore exhibit a rapidly changing normalized
fingerprint while transmitting very little power, resulting in few detected
photons for a fixed input power and integration time. The total throughput was
evaluated separately as
\begin{equation}
    T(\lambda)
    =
    \frac{\displaystyle\sum_{i=1}^{7}P_i(\lambda)}
    {P_{\mathrm{in}}(\lambda)}
    \label{eq:throughput}
\end{equation}
Because \(S_{\lambda}\) and \(\mathcal{I}_{\lambda}\) depend only on the
normalized port fractions, they do not quantify the total optical power
available for detection. Throughput was therefore evaluated separately for
the selected geometry after establishing time convergence. Only the
converged throughput was used as an absolute measure of device efficiency;
no efficiency conclusions were drawn from the short-duration screening
simulations.

The output power monitors were sampled from 720 to 770~nm in 1-nm increments. Perfectly
matched layers were applied on all six boundaries. No symmetry reduction was
used because the sequential-release fanout geometry is not mirror symmetric.
The initial sweeps used a common simulation time of 1.5~ps to provide a
computationally efficient comparison among candidate geometries. These runs
were therefore treated as a screening study, while the absolute response of
the selected geometry was determined using the longer convergence study
described in Sec.~\ref{subsec:time_convergence}.

\begin{figure}[htbp]
    \centering
    \includegraphics[width=0.8\linewidth]
    {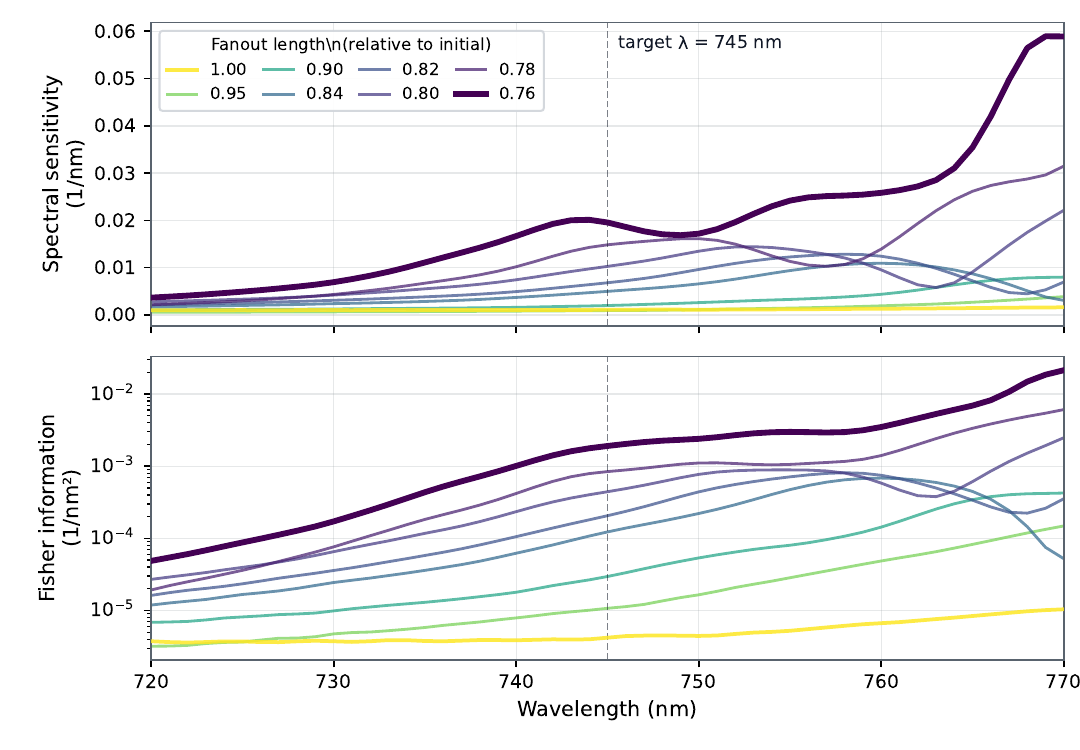}
    \caption{%
        Simulated spectral sensitivity of the normalized seven-port intensity
        vector (top) and corresponding multinomial Fisher information (bottom)
        for longitudinal scale factors \(s=1.00\)--0.76. The dashed vertical
        line indicates the 745-nm target wavelength. Shortening the multimode
        and sequential-fanout regions increases the wavelength-dependent
        redistribution of power and identifies \(s=0.76\) as the strongest
        candidate within the sampled range. These curves were obtained using
        the common 1.5-ps screening condition.
    }
    \label{fig:scale_sweep}
\end{figure}

\subsection{Longitudinal geometry optimization}
\label{subsec:longitudinal_optimization}

The principal geometry sweep varied a longitudinal scale factor \(s\) applied
to the shared multimode-interference region and to the longitudinal coordinates
of all seven fanout branches. The transverse geometry, material stack, branch
widths, final output pitch, source parameters, monitor cross sections, and
10-\(\mu\mathrm{m}\) straight output tails were held fixed. Eight scale factors
were evaluated:
\begin{equation}
    s =
    1.00,\ 0.95,\ 0.90,\ 0.84,\ 0.82,\ 0.80,\ 0.78,\ \mathrm{and}\ 0.76
\end{equation}
For the unscaled geometry, the multimode region and fanout ended at
\(x=45.0~\mu\mathrm{m}\) and \(x=110.0~\mu\mathrm{m}\), respectively. For the
shortest geometry, \(s=0.76\), these coordinates were reduced to
\(34.2~\mu\mathrm{m}\) and \(83.6~\mu\mathrm{m}\).

Figure~\ref{fig:scale_sweep} shows the spectral sensitivity and Fisher
information calculated across the 720--770-nm band. Shortening the device
moves a strong wavelength-dependent modal-beating feature toward the target
wavelength of 745~nm. At the target wavelength, the \(s=0.76\) geometry
produced the largest response among the sampled cases, with
\begin{equation}
    S_{\lambda}
    =
    1.95\times10^{-2}~\mathrm{nm}^{-1}
\end{equation}
and
\begin{equation}
    \mathcal{I}_{\lambda}
    =
    1.91\times10^{-3}~\mathrm{nm}^{-2}
\end{equation}
By comparison, the unscaled geometry produced
\(S_{\lambda}=1.08\times10^{-3}~\mathrm{nm}^{-1}\) and
\(\mathcal{I}_{\lambda}=4.20\times10^{-6}~\mathrm{nm}^{-2}\) at 745~nm.
The normalized-vector slope, Fisher information, and screened throughput
therefore all favored the shorter geometry near the target wavelength.

The response continued to improve down to the shortest geometry included in
the sweep. Consequently, \(s=0.76\) was selected as the best screened
candidate, but the sweep did not bracket a global longitudinal optimum.
Additional converged simulations below \(s=0.76\) would be required to
determine whether further shortening provides an additional benefit.

\subsection{Time-domain convergence}
\label{subsec:time_convergence}

Following the longitudinal sweep, the selected \(s=0.76\) layout was evaluated
with \(0.5~\mu\mathrm{m}\)-wide output waveguides in a dedicated
time-convergence study. The geometry, material definitions, broadband source,
and monitor locations were held fixed while the FDTD simulation time was
increased through
\begin{equation}
    t_{\mathrm{sim}}
    =
    3,\ 4,\ 5,\ 6,\ 8,\ 10,\ 12,\ \mathrm{and}\ 14~\mathrm{ps}
\end{equation}
The output was sampled at 743, 745, and 747~nm so that convergence of both the
throughput and the local wavelength-dependent port redistribution could be
tested.

Convergence was declared only when, relative to the preceding simulation, all
three throughput values and all 21 normalized port fractions changed by no
more than 2\%. This criterion is more stringent than examining the throughput
at 745~nm alone because small residual changes in a weak port can affect the
calculated spectral sensitivity and Fisher information.

\begin{figure}[h]
    \centering
    \includegraphics[width=0.82\linewidth]
    {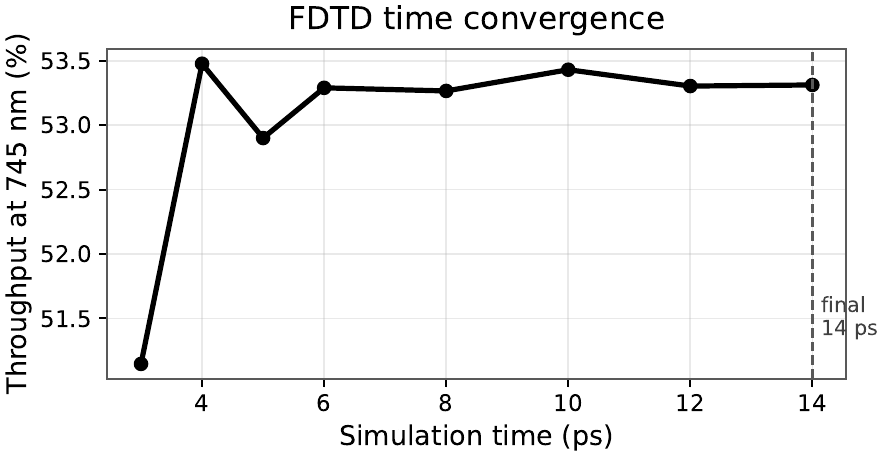}
    \caption{%
        Throughput at 745~nm as a function of FDTD simulation time for the
        selected \(s=0.76\) geometry with \(0.5~\mu\mathrm{m}\)-wide output
        waveguides. Although the scalar throughput stabilizes near 53.3\% at
        shorter simulation times, the 14-ps result is the first to satisfy the
        2\% convergence requirement simultaneously for all three sampled
        throughput values and all 21 wavelength-port fractions.
    }
    \label{fig:time_convergence}
\end{figure}

As shown in Fig.~\ref{fig:time_convergence}, the 745-nm throughput approached
approximately 53.3\% after the initial transient. The complete convergence
criterion first passed for the 14-ps simulation relative to the 12-ps result.
The maximum relative change among the three throughput values was 0.0413\%,
and the maximum relative change among the 21 normalized port fractions was
1.268\%. The largest absolute change in an individual normalized port
fraction was 0.0715 percentage points.

The converged throughputs were 49.57\%, 53.31\%, and 61.67\% at 743, 745, and
747~nm, respectively. The local derivative at 745~nm was evaluated using the
centered finite difference
\begin{equation}
    \left.
    \frac{\partial\mathbf{p}}{\partial\lambda}
    \right|_{745\,\mathrm{nm}}
    \approx
    \frac{
        \mathbf{p}(747\,\mathrm{nm})
        -
        \mathbf{p}(743\,\mathrm{nm})
    }{4\,\mathrm{nm}}.
    \label{eq:centered_difference}
\end{equation}
The converged response gave
\(S_{\lambda}=8.42\times10^{-2}~\mathrm{nm}^{-1}\) and
\(\mathcal{I}_{\lambda}=6.07\times10^{-2}~\mathrm{nm}^{-2}\). Between the
12- and 14-ps simulations, these quantities changed by only 0.040\% and
0.054\%, respectively. A simulation duration of 14~ps was therefore adopted
for the final local spectral calculations.

\subsection{Scope and limitations of the optimization}
\label{subsec:optimization_limitations}

The geometry sweep and convergence study serve distinct purposes. The
longitudinal sweep provides a computationally efficient comparison among
candidate geometries and identifies the direction in parameter space that
increases wavelength discrimination. The 14-ps convergence study instead
establishes the quantitative response of the selected design. Because the
initial 1.5-ps simulations were intended only for screening and were not
temporally converged, their absolute metric values should not be compared
directly with the converged results. The screening runs are therefore used to
identify relative geometric trends, while absolute device performance is
reported only from the converged simulation.

Furthermore, convergence of the selected geometry does not by itself
demonstrate that the ranking of every longitudinal scale factor remains
unchanged at 14~ps. The present results therefore support \(s=0.76\) as a
promising design candidate rather than a confirmed global optimum. A compact,
fully converged refinement sweep below and around \(s=0.76\), together with
mesh-convergence and fabrication-tolerance studies, would provide the final
validation of the optimized geometry.

\section{Results}
\label{sec:results}

\subsection{Wavelength-dependent output fingerprints}
\label{subsec:output_fingerprints}

\begin{figure}[htbp]
    \centering
    \includegraphics[width=1\linewidth]
    {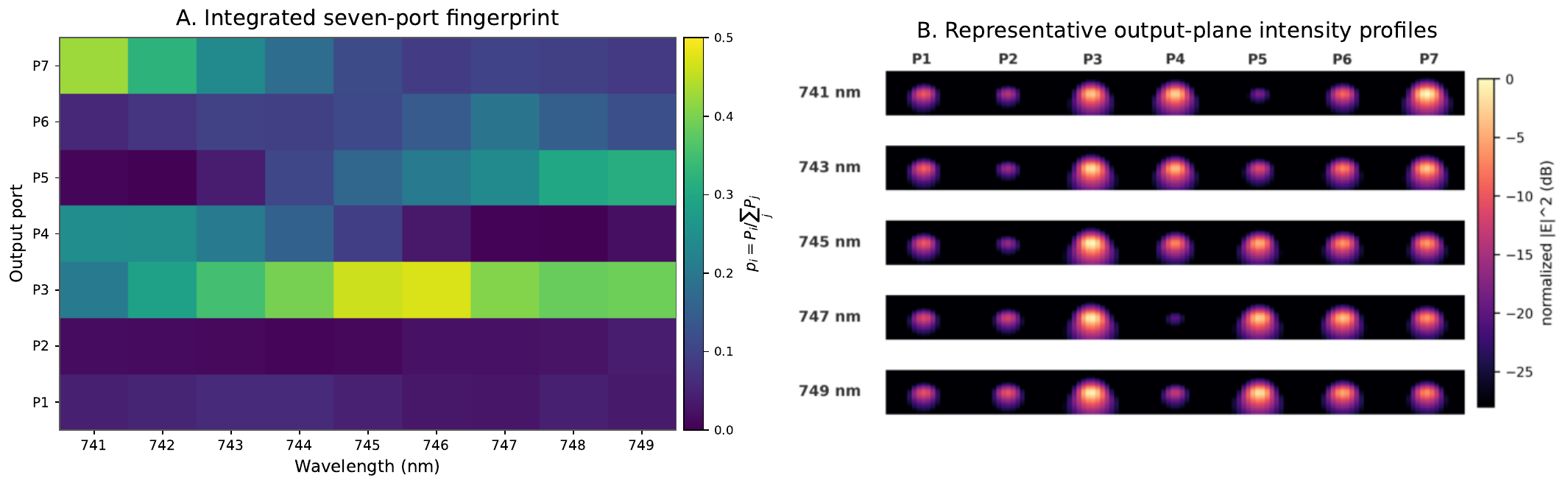}
    \caption{%
        Wavelength-dependent response of the refined \(s=0.76\) design from
        the converged 14-ps simulation. (a) Normalized power fraction
        \(p_i=P_i/\sum_jP_j\) in each of the seven output ports from 741 to
        749~nm. (b) Representative output-plane intensity profiles at 741,
        743, 745, 747, and 749~nm, displayed using a common decibel scale.
        The changing relative port intensities form a wavelength-dependent
        spatial fingerprint.
    }
    \label{fig:refined_fingerprint}
\end{figure}

The refined \(s=0.76\) design produced a distinct seven-port intensity fingerprint at each
sampled wavelength. Figure~\ref{fig:refined_fingerprint} shows the normalized
port fractions obtained from the converged 14-ps simulation over the
741--749-nm band. The total simulated throughput remained between 49.6\% and
67.8\% across this interval and was 53.3\% at the 745-nm design wavelength.
Thus, the wavelength-dependent redistribution was obtained while retaining a
substantial fraction of the input power.

At 741~nm, the output was concentrated primarily in P7, P4, and P3, which
carried 42.8\%, 24.5\%, and 20.5\% of the transmitted power, respectively.
At 745~nm, P3 became the dominant output with 46.2\%, while P5, P7, P6, and
P4 carried 16.5\%, 11.3\%, 10.9\%, and 9.3\%, respectively. By 749~nm, the
response was concentrated mainly in P3 and P5, with normalized fractions of
38.9\% and 31.1\%. An especially pronounced feature occurs near 747~nm,
where the P4 fraction falls to 0.44\% while P3, P5, and P6 carry 40.7\%,
24.0\%, and 19.3\%, respectively.

These changes demonstrate that wavelength is encoded in the relative
intensities of several output channels rather than in the transmission of a
single port. The output-plane field profiles in
Fig.~\ref{fig:refined_fingerprint}(b) show the same redistribution directly:
the brightest output spot moves among the ports as the wavelength is varied,
while the fields remain spatially localized at the output waveguides.

\subsection{Spectral sensitivity and Fisher information}
\label{subsec:refined_spectral_metrics}

\begin{figure}[htbp]
    \centering
    \includegraphics[width=0.8\linewidth]
    {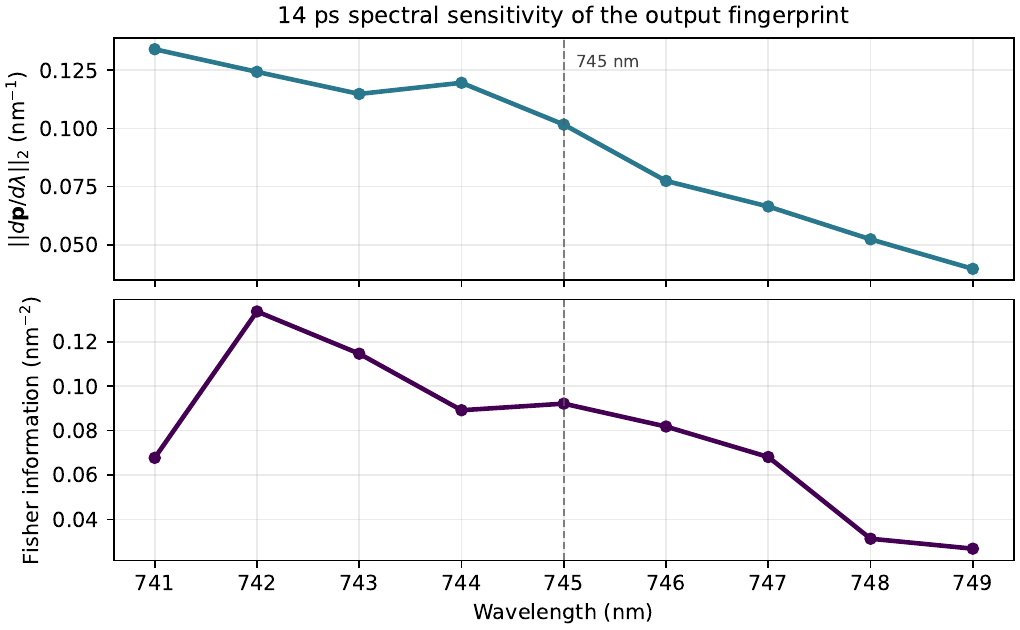}
    \caption{%
        Spectral sensitivity of the normalized seven-port output vector
        (top) and multinomial Fisher information per detected photon
        (bottom) for the finalized design, calculated from the converged
        14-ps simulation. The dashed line marks the 745-nm design wavelength,
        where \(S_{\lambda}=0.102~\mathrm{nm}^{-1}\) and
        \(\mathcal{I}_{\lambda}=0.0921~\mathrm{nm}^{-2}\). Interior
        wavelengths were evaluated using centered finite differences; the
        values at 741 and 749~nm were evaluated using one-sided differences.
    }
    \label{fig:refined_metrics}
\end{figure}

The strong port-to-port redistribution produces a correspondingly large
spectral sensitivity, as shown in Fig.~\ref{fig:refined_metrics}. The
normalized-vector sensitivity was \(0.134~\mathrm{nm}^{-1}\) at 741~nm and
\(0.102~\mathrm{nm}^{-1}\) at the 745-nm design wavelength. It decreased
toward the red side of the simulated interval but remained
\(0.0665~\mathrm{nm}^{-1}\) at 747~nm and
\(0.0397~\mathrm{nm}^{-1}\) at 749~nm. The response is therefore strongest
on the blue side of the selected band while remaining wavelength-dependent
throughout the full 8-nm interval. Because 741 and 749~nm are the boundaries
of the simulated interval, their derivatives were evaluated using one-sided
finite differences; the interior values, including that at 745~nm, were
evaluated using centered differences.

At 745~nm, the local derivative of the normalized output vector was
\begin{equation}
    \left.
    \frac{\partial\mathbf{p}}{\partial\lambda}
    \right|_{745\,\mathrm{nm}}
    =
    \begin{bmatrix}
        -0.0135 &
         0.00883 &
         0.0371 &
        -0.0609 &
         0.0488 &
         0.0245 &
        -0.0449
    \end{bmatrix}^{\mathrm{T}}
    \mathrm{nm}^{-1}
    \label{eq:simulated_port_derivative}
\end{equation}
The dominant changes correspond to P4 decreasing by approximately
6.09 percentage points per nanometer, P5 increasing by 4.88 percentage
points per nanometer, P7 decreasing by 4.49 percentage points per nanometer,
and P3 increasing by 3.71 percentage points per nanometer. The wavelength
information is therefore distributed across multiple ports, reducing
dependence on any single detector channel.

The multinomial Fisher information reached its maximum within the sampled
interval at 742~nm, where it was \(0.134~\mathrm{nm}^{-2}\), and was
\(0.0921~\mathrm{nm}^{-2}\) at 745~nm. After including the simulated
throughput, the throughput-weighted Fisher information at the design
wavelength was \(0.0491~\mathrm{nm}^{-2}\). The combination of finite
throughput and a rapidly varying normalized output vector indicates that the
device does not obtain its spectral response solely by operating near a
transmission null.

The curves in Fig.~\ref{fig:refined_metrics} differ substantially in shape
and magnitude from those obtained during the initial geometry screening.
This difference results primarily from the distinct purposes and numerical
conditions of the two studies. The geometry sweep used a fixed 1.5-ps
simulation window to compare many candidate geometries at manageable
computational cost, whereas the results above were obtained from the
converged 14-ps response of the finalized design. Subsequent diagnostic
testing confirmed that the 1.5-ps output fingerprint was time truncated: in
an otherwise identical 745-nm control, increasing the simulation duration
from 1.5 to 3~ps changed the collected throughput from approximately 0.67\%
to 51.1\% and substantially altered the normalized seven-port response.

Time truncation affects the plotted metrics more strongly than it affects
the individual port fractions because both metrics are calculated from
wavelength derivatives. The Fisher information is particularly sensitive
because each squared derivative is additionally weighted by \(1/p_i\), so
changes in relatively weak ports can produce large changes in its magnitude
and spectral shape. The screening and converged curves use the same metric
definitions, but the former describe a transient response and the latter
describe the settled output fingerprint. Their absolute values therefore
should not be expected to coincide.

The initial sweep is consequently interpreted as a comparative
down-selection step that identified \(s=0.76\) as a promising candidate,
while the quantitative performance claims are based exclusively on the
converged calculation. Because convergence was established for the selected
design rather than for every geometry in the original sweep, the screening
does not by itself demonstrate that \(s=0.76\) is a global optimum. A
converged comparison with neighboring scale factors would be required to
confirm the final geometry ranking.

\subsection{Field evolution through the multimode fanout}
\label{subsec:field_evolution}

\begin{figure}[htbp]
    \centering
    \includegraphics[width=0.8\linewidth]
    {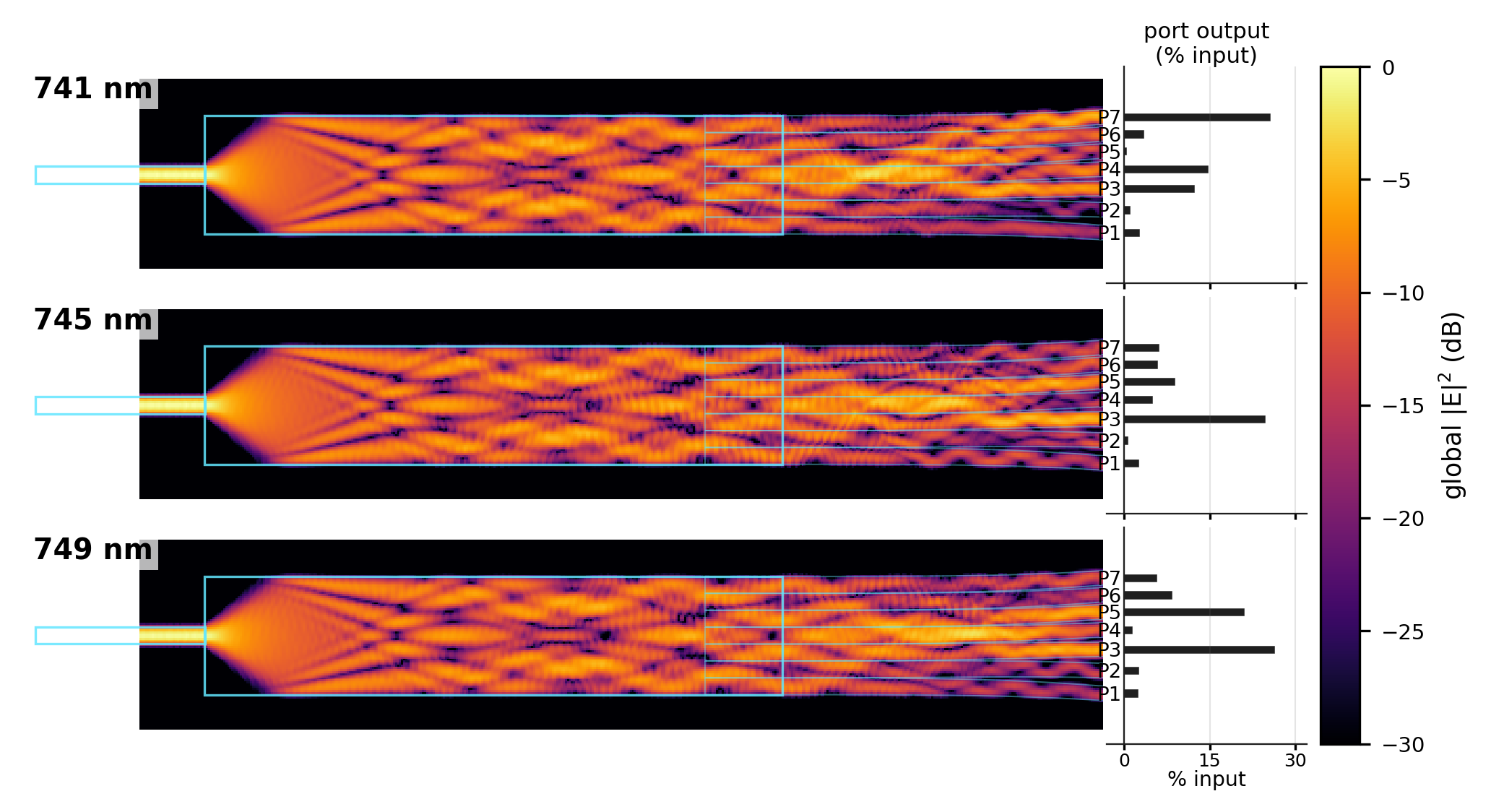}
    \caption{%
        Representative longitudinal intensity distributions at 741, 745,
        and 749~nm for the refined \(s=0.76\) geometry. The field maps,
        extracted from the z-slice-bearing screening model and shown on a
        common global decibel scale, illustrate the wavelength-dependent
        multimode interference pattern and its evolution through the
        sequential fanout. The adjacent bars give the converged 14-ps power
        delivered to each output port as a percentage of the input power.
    }
    \label{fig:field_evolution}
\end{figure}

The longitudinal field distributions provide a physical interpretation of
the output fingerprints. As shown in Fig.~\ref{fig:field_evolution}, the
input field excites a multimode interference pattern in the shared region.
Small changes in wavelength alter the accumulated relative phases of these
modes before they enter the sequential fanout. The fanout then samples
different portions of the interference pattern and transfers them into the
seven separated output waveguides.

At 741~nm, the interference pattern preferentially directs power toward the
upper outputs, producing strong P7, P4, and P3 signals. At 745~nm, the field
distribution shifts toward P3 and P5, with P3 carrying the largest fraction
of the transmitted power. At 749~nm, P5 grows further while P3 remains
strong. The wavelength response is consequently generated by coherent
multimode evolution throughout the shared region and fanout rather than by a
single localized resonance. This distributed mechanism produces smooth but
readily distinguishable output fingerprints over the simulated wavelength
range.

\subsection{Summary of refined-design performance}
\label{subsec:refined_results_summary}

The refined simulation demonstrates three properties required for
intensity-based wavelength discrimination. First, the device maintains
49.6--67.8\% simulated throughput over the 741--749-nm interval. Second, the
normalized output pattern changes strongly with wavelength and remains
nondegenerate across the seven ports. Third, the target-wavelength response
combines a spectral sensitivity of \(0.102~\mathrm{nm}^{-1}\), a multinomial
Fisher information of \(0.0921~\mathrm{nm}^{-2}\), and a throughput-weighted
Fisher information of \(0.0491~\mathrm{nm}^{-2}\). Together, these results
show that the refined fanout converts small wavelength changes into
measurable, spatially distributed intensity changes without requiring
spectrally resolved detection at the outputs.

\section{Conclusion}
\label{sec:conclusion}

We investigated a seven-port multimode-interference fanout that converts
small wavelength changes into measurable changes in a spatial intensity
fingerprint. Three-dimensional FDTD simulations were used first to screen
the longitudinal geometry and then to evaluate the selected design under
time-converged conditions. The screening study showed that shortening the
multimode and fanout regions increased the wavelength dependence near the
745-nm target, leading to the selection of the \(s=0.76\) geometry for
further analysis. Because these initial simulations used a time-truncated
1.5-ps window, they were treated as a comparative down-selection step rather
than as a quantitative prediction of the final performance.

The finalized design was evaluated over 741--749~nm using a 14-ps simulation
duration established through a dedicated convergence study. Across this
interval, the device maintained a simulated throughput of 49.6--67.8\% and
produced distinct wavelength-dependent distributions among the seven output
ports. At the 745-nm design wavelength, the throughput was 53.3\%, the
normalized-vector spectral sensitivity was
\(0.102~\mathrm{nm}^{-1}\), and the multinomial Fisher information was
\(0.0921~\mathrm{nm}^{-2}\). Including the simulated throughput gave a
throughput-weighted Fisher information of
\(0.0491~\mathrm{nm}^{-2}\). The wavelength information was distributed
across several ports rather than being associated with a single output
channel or a transmission null. The simulated field distributions indicate
that this response arises from wavelength-dependent multimode evolution
followed by spatial sampling through the sequential fanout.

These results demonstrate the simulated feasibility of using a compact
multiport intensity response for wavelength discrimination without
spectrally resolving each output. The present study nevertheless establishes
a promising design candidate rather than a globally optimized or
experimentally validated device. A first priority for future work is a
converged geometry refinement around and below \(s=0.76\). Simulating several
neighboring scale factors with the final time and mesh settings will determine
whether the screening ranking is retained and will bracket the longitudinal
optimum. Additional numerical studies should examine spatial-mesh
convergence, guided-mode coupling at each output, and the wavelength
dependence of the insertion-loss budget.

Fabrication-tolerance analysis will also be required to quantify sensitivity
to waveguide width, film thickness, refractive index, sidewall angle,
alignment, and surface roughness. Because the device is intended to operate
from a calibrated multiport fingerprint, such perturbations need not leave
the response unchanged, but they must preserve a sufficiently strong and
nondegenerate wavelength mapping. Extending the simulations to a broader and
more finely sampled spectral interval will help identify possible fingerprint
degeneracies and establish the usable measurement bandwidth.

Finally, the optical model should be combined with a realistic detector and
estimation model incorporating photon-counting statistics, detector noise,
channel-to-channel calibration errors, and finite optical bandwidth.
Reconstruction algorithms can then be evaluated against the simulated
Fisher information for specified photon budgets. Fabrication and experimental calibration of the device will provide
the ultimate test of throughput, repeatability, environmental stability, and
achievable wavelength precision.
\acknowledgments 

A.P. would like to thank Rachel Morgan and Johnathon Lin for valuable discussions. This work was funded by the Kavli Foundation. The authors gratefully acknowledge its support of this research. 

\bibliography{report} 
\bibliographystyle{spiebib} 

\end{document}